\documentclass[epj]{svjour}
\usepackage{graphics}
\begin{document}
\title{Studies of Inclusive Jet Production in ep Interactions at HERA}

\author{M\'ONICA V\'AZQUEZ
\inst{ }
}                     
\institute{Departamento de F\'{\i}sica Te\'orica, Universidad Aut\'onoma de Madrid, Cantoblanco, E-28
049
Madrid, Spain\\
e-mail: monicava@mail.desy.de}
\date{Received: 15 August, 2003}
%
\abstract{Inclusive jet production in neutral and charged current deep inelastic scattering, in photoproduction and the transition region has been studied with the ZEUS and H1 detectors at HERA. The measurements have been compared 
to next-to-leading-order QCD calculations which are used to make determinations of the strong coupling constant.
} 
\maketitle
\section{Introduction \label{intro}}
\vspace*{-0.2cm}
The HERA collider provides a unique laboratory for the study of the hadronic final state, in terms of the jet production rates in $ep$ collisions. Depending on the virtuality, $Q^2$, of the exchanged gauge boson two different regimes are studied; in the case of large $Q^2$ ($Q^2\gg 1$ GeV$^2$) it is referred to as deep inelastic scattering (DIS) and in the case the exchanged photon is quasi-real ($Q^2\sim 0$) it is referred to as photoproduction. Jet data are very precise at high transverse energy in both the DIS and photoproduction regimes, where the experimental uncertainties and non-perturbative effects are small. This allows precision tests of perturbative QCD to be perform and further constraints to be placed on the proton and photon parton distribution functions (PDFs). 
Most of the subprocesses studied have leading contributions proportional to the strong coupling, $\alpha_s$, which enables the extraction of QCD parameters. 
The low $Q^2$ transition region ($5<Q^2<100$~GeV$^2$) between photoproduction and DIS can also be studied, a region where non-perturbative effects and the theoretical uncertainties become more important.

Inclusive jet cross sections in photoproduction, neutral current (NC) and charged current (CC) DIS and the transition region have been measured using the data collected with the ZEUS and H1 detectors at HERA. The measurements are compared to next-to-leading-order (NLO) QCD calculations.

\vspace*{-0.5cm}
\section{Inclusive Jets in Photoproduction \label{sec:ph}}
\vspace*{-0.25cm}
In photoproduction at HERA the photon has low virtuality, $Q^2\sim 10^{-3}$~GeV$^2$, and the high transverse energy of the jets provides the hard scale. In the study of inclusive jet production the phase space of the second jet is not restricted. This reduces the theoretical uncertainties in the NLO QCD predictions although the information on the event kinematics is reduced.

\begin{figure}[t]
\vspace*{-0.3cm}  
\begin{center}
\resizebox{0.47\textwidth}{!}{%
\includegraphics{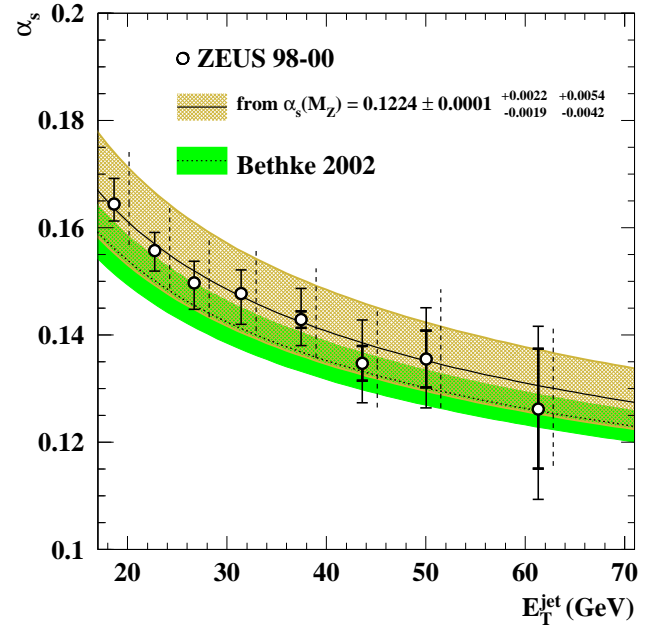}
}
\end{center}
\vspace*{-0.7cm}       
\caption{The $\alpha_s(E_T^{\rm jet})$ values determined from a QCD fit of the measured $d\sigma/dE_T^{\rm jet}$ as a function of  $E_T^{\rm jet}$.}
\label{fig:runalphas}       
\end{figure}

\vspace*{-0.6cm}
The ZEUS collaboration has measured the inclusive jet cross section in photoproduction~\cite{ph_zeus} as a function of the transverse energy of the jets, $E_T^{\rm jet}$. NLO QCD calculations~\cite{klasen} agree well over 5 orders of magnitude with the measured cross section. A NLO QCD $\chi^2$-fit to the measurement has allowed the extraction of values of $\alpha_s$ for different $E_T^{\rm jet}$ values which are presented in Figure~\ref{fig:runalphas}. The running of $\alpha_s$ is seen in a single measurement. A fit to all the  $E_T^{\rm jet}$ gives a value of $\alpha_s(M_Z)$ of
\begin{displaymath}
 \alpha_s(M_Z)=0.1224\pm 0.0001(stat.) ^{+0.0022}_{-0.0019}(exp.)  ^{+0.0054}_{-0.0042} (th.), \nonumber
\end{displaymath}
which is consistent with a recent determination of Bethke~\cite{bethke}.

\begin{figure}[t]
\begin{center}
\resizebox{0.5\textwidth}{!}{%
\includegraphics{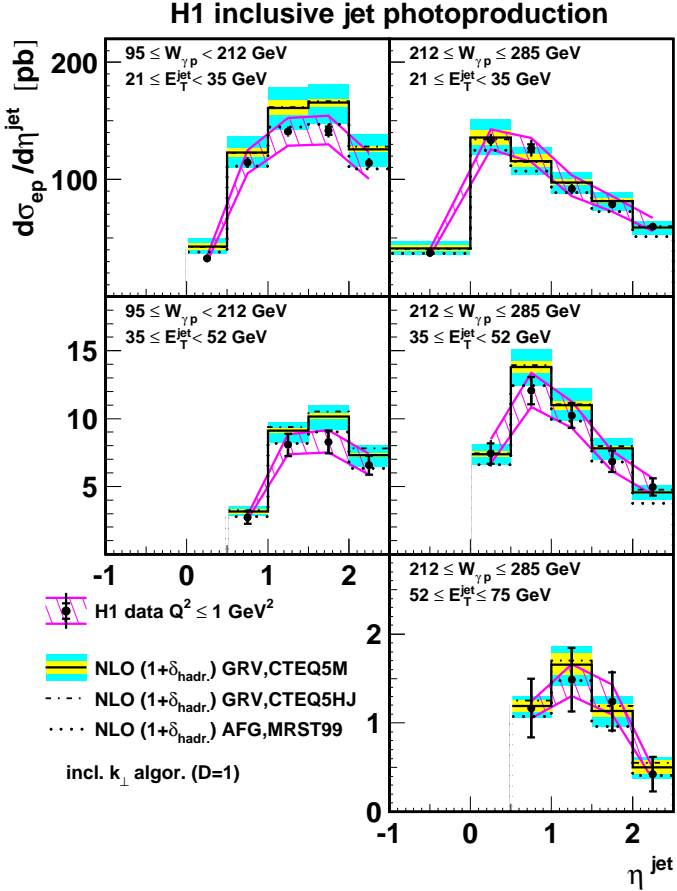}
}
\end{center}
\vspace*{-0.3cm}       
\caption{Inclusive jet cross section in photoproduction as a function of $\eta^{\rm jet}$ in different $E_T^{\rm jet}$ and $W_{\gamma p}$ regions.}
\label{fig:phpeta}       
\end{figure}

The H1 collaboration has measured inclusive jet cross sections as a function of the jet pseudorapidity\cite{ph_h1}, $\eta^{\rm jet}$, in different regions of $E_T^{\rm jet}$ and hadronic center-of-mass energy, $W_{\gamma p}$ shown in Figure~\ref{fig:phpeta}. The NLO QCD calculations~\cite{frixione} using different proton and photon PDFs agree well with the data. 

\vspace*{-0.3cm}
\section{Inclusive Jets in Deep Inelastic Scattering \label{sec:dis}}
\vspace*{-0.2cm}
Jet production in DIS proceeds up to leading order in the strong coupling constant via the Born ($\gamma /Z/W \, q \rightarrow q$), the QCD-Compton ($\gamma /Z/W \, q \rightarrow qg$) and boson-gluon fusion ($\gamma /Z/W \, g \rightarrow q\overline{q}$) processes. In the Breit frame (struck parton and gauge-boson collide on-head) the Born contribution is supressed and processes directly porportional to $\alpha_s$ are studied.

\begin{figure}[t]
\begin{center}
\resizebox{0.5\textwidth}{!}{%
\includegraphics{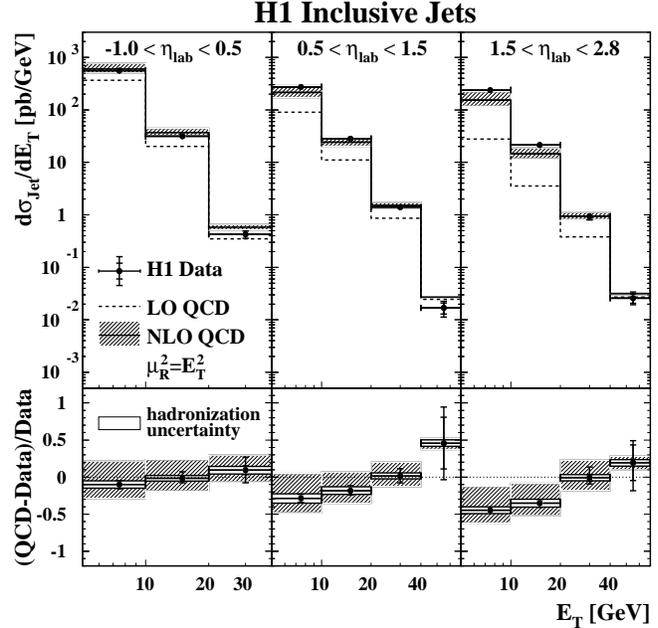}
}
\end{center}
\vspace*{-0.6cm}       
\caption{Inclusive jet cross section in NC DIS, in the kinematic region $5 < Q^2 < 100$ GeV$^2$, as a function of $E_T^{\rm jet}$ in the Breit frame in different $\eta^{\rm jet}$ regions.}
\label{fig:ncet}       
\end{figure}

At large virtualities, $Q^2\geq 125$ GeV$^2$, NLO QCD calculations have been seen to describe well the the measured jet cross sections in NC DIS. At low virtualities discrepancies were observed previously and are studied in more detail by the H1 collaboration~\cite{nc_h1}. The measured inclusive jet cross section as a function of $E_T^{\rm jet}$ in the Breit frame is shown in Fig.~\ref{fig:ncet} for the backward, central and forward regions. The NLO QCD prediction~\cite{seymour} agrees well with the data in the backward and central regions and is below the data for $E_T^{\rm jet}<20$ GeV in the forward region, where the LO to NLO corrections and the renormalization scale uncertainties are largest.   

\begin{figure}[b]
\vspace*{-1.2cm}
\begin{center}
\resizebox{0.4\textwidth}{!}{%
\includegraphics{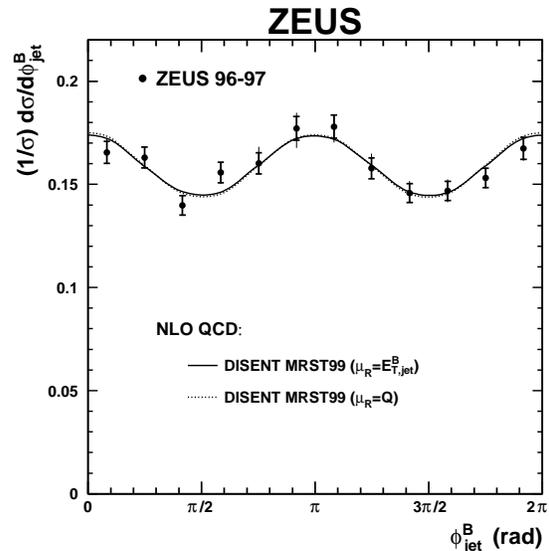}
}
\end{center}
\vspace*{-0.6cm}       
\caption{Normalised studied jet cross section in NC DIS as a function of the azimuthal angle $\phi^B_{\rm jet}$ in the Breit frame.}
\label{fig:az}       
\end{figure}

The ZEUS collaboration has measured the azimuthal distribution of jets produced in high-$Q^2$ ($Q^2\geq 125$ GeV$^2$) NC DIS~\cite{nc_zeus}. The normalised differential cross section as function of the azimuthal angle in the Breit frame, $\phi^{B}_{jet}$, between the lepton scattering plane and the plane defined by the jet and the incoming proton direction has been measured and is shown in Fig.~\ref{fig:az}. The distribution is enhanced at $\phi^{B}_{jet}=0$ and $\phi^{B}_{jet}=\pi$ and is well described by the NLO QCD calculation~\cite{seymour}.


\clearpage

The dependence of the azimuthal asymmetry with $Q^2$ has been also measured and the asymmetry is seen to decrease as $Q^2$ increases. This behaviour is also observed in the NLO QCD calculation and can be attributed to the progressive decline of the boson-gluon fusion process as $Q^2$ increases.

The ZEUS collaboration has measured inclusive jet cross sections in CC DIS~\cite{cc_zeus}, which allows to test flavour changing electroweak theory and QCD in one type of events. Inclusive jet cross sections in the laboratory frame as a function of $E_T^{\rm jet}$ and $\eta^{\rm jet}$ have been measured and are in good agreement with NLO QCD calculations~\cite{zeppenfeld}. 

\begin{figure}[b]
\begin{center}
\resizebox{0.5\textwidth}{!}{%
\includegraphics{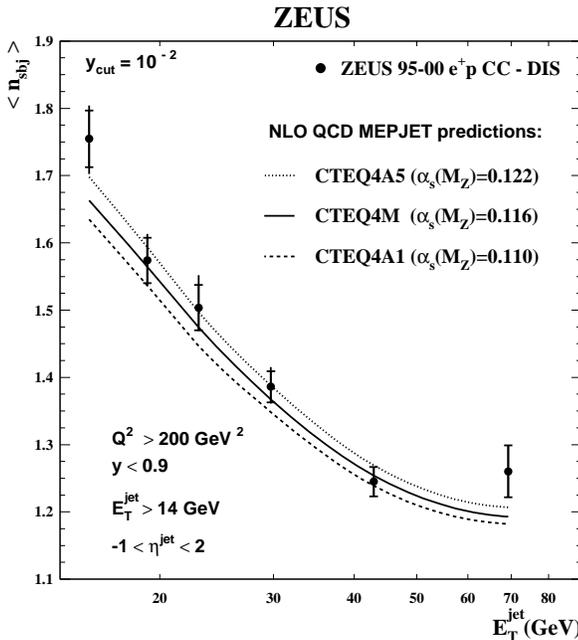}
}
\end{center}
\vspace*{-0.6cm}       
\caption{Mean subjet multiplicity of a inclusive jet sample in CC DIS as a function of $E_T^{\rm jet}$.}
\label{fig:subjet}       
\end{figure}

The internal structure of the inclusive jet sample in CC DIS has been studied in terms of the mean subjet multiplicity~\cite{su}, $\langle n_{sbj} \rangle$. The measured $\langle n_{sbj} \rangle$ as a function of $E_T^{\rm jet}$ for a fixed resolution scale $y_{cut}=10^{-2}$ decreases as $E_T^{\rm jet}$ increases (see Fig.~\ref{fig:subjet}). The measured $\langle n_{sbj} \rangle$ is compared to NLO QCD calculations and $\alpha_s$ has been determined as a NLO QCD $\chi^2$-fit to the data. The extracted value from $\langle n_{sbj}\rangle$ for jets with $E_T^{jet}>25$~GeV is
\begin{displaymath}
 \alpha_s(M_Z)=0.1202\pm 0.0052(stat.) ^{+0.0060}_{-0.0019}(exp.)  ^{+0.0065}_{-0.0053} (th.),
\end{displaymath}
which is consistent with other recent determinations.

The mean subjet multiplicity as a function of $Q^2$ has been compared to a similar measurement in NC DIS and is found to be consistent. The QCD radiation does not depend on the gauge boson exchanged, a $W$ in CC DIS or a $\gamma$ in NC DIS.

\section{Summary and Outlook \label{summ}}
\vspace*{-0.2cm}
HERA is producing a wealth of precision jet data at high transverse energies in neutral current and charged current deep inelastic scattering, photoproduction and the transition region.

At high jet transverse energies in deep inelastic scattering and photoproduction, the theoretical uncertainties in the inclusive jet cross section calculations are small and the predictions are able to reproduce cross sections over many orders of magnitude. In the transition region of lower values of the virtuality $Q^2$ of the exchange boson, the theoretical uncertainties are dominant and further theoretical developments would be desirable.

Many extractions of the QCD coupling constant $\alpha_s$ are performed with a precision competitive with the world average. The running behaviour of $\alpha_s$ has been observed using the photoproduction jet data and the theoretical uncertainty dominates the precision of $\alpha_s$. 

The experimental precision and coverage of the data is good and it is time to include the HERA jet data in global PDF fits. This would allow better constraints to be placed on the gluon content of the proton and the hadron-like structure of the photon.



\begin{thebibliography}{}
\vspace*{-0.2cm}
%
%
\bibitem{ph_zeus}
ZEUS Collaboration, 
Phys. Lett. \textbf{B 560} (2003) 7.

\bibitem{ph_h1}
H1 Collaboration, 
hep-ex/0302034.

\bibitem{nc_h1}
H1 Collaboration, 
 Phys. Lett. \textbf{B 542} (2002) 193.

\bibitem{nc_zeus}
ZEUS Collaboration, 
Phys. Lett. \textbf{B 551} (2003) 226.

\bibitem{cc_zeus}
ZEUS Collaboration, 
hep-ex/0306018.

\bibitem{klasen}
M. Klasen et al., Eur. Phys. J. direct{\bf C 1} (1998) 1.

\bibitem{bethke}
S.~Bethke, hep-ex/0211012.

\bibitem{frixione}
S.~Frixione and G.~Ridolfi, Nucl. Phys. {\bf B 507} (1997) 315.

\bibitem{seymour}
S. Catani and M.H. Seymour, Nucl. Phys. {\bf B 485} (1997) 291. Erratum in Nucl. Phys. {\bf B 510} (1998) 503.

\bibitem{zeppenfeld}
E.~Mirkes and D.~Zeppenfeld, Phys. Lett. {\bf B 380} (1996) 205.

\bibitem{su} S.~Catani et al., Nucl. Phys. {\bf B 377} (1992) 445 and Nucl. Phys. {\bf B 383} (1992) 419;\\
 M.H. Seymour, Nucl. Phys. {\bf  B 421} (1994) 545 and Phys. Rev. Lett. {\bf B 378} (1996) 279.


\end{thebibliography}
\end{document}